\documentclass{emulateapj}

\usepackage{amsmath}
\usepackage{mathrsfs}
\usepackage{color}

\newcommand{\simgt}{\lower.5ex\hbox{$\; \buildrel > \over \sim \;$}}
\newcommand{\simlt}{\lower.5ex\hbox{$\; \buildrel < \over \sim \;$}}

\slugcomment{To appear in The Astrophysical Journal}
\makeatletter

\begin{document}
  
\shorttitle{
  A clipping method to mitigate the photo-$z$ errors
}
\title{
  A Clipping Method to Mitigate the
  Impact of Catastrophic Photometric Redshift Errors on Weak Lensing Tomography
}

\author{
  Atsushi J. Nishizawa\altaffilmark{1},
  Masahiro Takada\altaffilmark{2},
  Takashi Hamana\altaffilmark{3},
  Hisanori Furusawa\altaffilmark{3}
}

\altaffiltext{1}{ 
  Astronomical Institute, Tohoku University Aramaki Aobaku Sendai 
  980-8578, Japan
} 
\altaffiltext{2}{
  Institute for the Physics and Mathematics of the Universe (IPMU), 
  the University of Tokyo, 5-1-5 Kashiwa-no-Ha, Kashiwa City, 
  Chiba 277-8582, Japan
}
\altaffiltext{3}{
  National Astronomical Observatory of Japan, 2-21-1
  Osawa Mitaka City Tokyo 181-8588, Japan
}

\shortauthors{Nishizawa et al.}
\email{nishizawa@astr.tohoku.ac.jp}

%%%%%%%%%%%%%%%%%%%%%%%%%%%%%%%%%%%%%%%%%%%%%%%%%%%%%%%%%%%%%%%%%%%%%%%%%%%%%%%
\begin{abstract} 
%%%%%%%%%%%%%%%%%%%%%%%%%%%%%%%%%%%%%%%%%%%%%%%%%%%%%%%%%%%%%%%%%%%%%%%%%%%%%%%
We use the mock catalog of galaxies, constructed based on the COSMOS
galaxy catalog including information on photometric redshifts
(photo-$z$) and SED types of galaxies, in order to study how to define
a galaxy subsample suitable for weak lensing tomography feasible with
optical (and NIR) multi-band data.  Since most of useful cosmological
information arises from the sample variance limited regime for
upcoming lensing surveys, a suitable subsample can be obtained by
discarding a large fraction of galaxies that have less reliable
photo-$z$ estimations. We develop a method to efficiently identify
photo-$z$ outliers by monitoring the width of posterior likelihood
function of redshift estimation for each galaxies.  This clipping
method may allow to obtain {\em clean} tomographic redshift bins (here
three bins considered) that have almost no overlaps between different
bins, by discarding more than $\sim 70\%$ galaxies of ill-defined
photo-$z$'s corresponding to the number densities of remaining
galaxies less than $\sim 20$ per square arcminutes for a Subaru-type
deep survey.  Restricting the ranges of magnitudes and redshifts
and/or adding near infrared data help obtain a cleaner redshift
binning.  By using the Fisher information matrix formalism, we
propagate photo-$z$ errors into biases in the dark energy equation of
state parameter $w$. We found that, by discarding most of ill-defined
photo-$z$ galaxies, the bias in $w$ can be reduced to the level
comparable to the marginalized statistical error, however, the
residual, small systematic bias remains due to asymmetric scatters
around the relation between photometric and true redshifts.  We also
use the mock catalog to estimate the cumulative signal-to-noise
($S/N$) ratios for measuring the angular cross-correlations of
galaxies between finner photo-$z$ bins, finding the higher $S/N$
values for photo-$z$ bins including photo-$z$ outliers.

\end{abstract}


%%%%%%%%%%%%%%%%%%%%%%%%%%%%%%%%%%%%%%%%%%%%%%%%%%%%%%%%%%%%%%%%%%%%%%%%%%%%%%%
\section{Introduction}
\label{sec:intro}
%%%%%%%%%%%%%%%%%%%%%%%%%%%%%%%%%%%%%%%%%%%%%%%%%%%%%%%%%%%%%%%%%%%%%%%%%%%%%%%

The bending of light by mass, gravitational lensing, causes images of
distant galaxies to be distorted \citep[e.g.][for a thorough
review]{BartelmannSchneider01}. These sheared source galaxies are
mostly too weakly distorted to measure the effect in individual
galaxies, but requires surveys containing at least millions of
galaxies to detect the signal in a statistical way
\citep[e.g. see][for the latest measurement result]{Fuetal:2008}.
This cosmic shear is now recognized as one of the most promising probe
that allows a direct reconstruction of the dark matter distribution as
well as to constrain the properties of dark energy or to test the
theory of gravity on cosmological scales \citep[e.g.][for recent
reviews]{HoekstraJain:2008,Masseyetal:2010,Huterer:2010}.  In
particular, by adding redshift information of source galaxies the
lensing geometrical information as well as the redshift evolution of
dark matter clustering can be inferred, thereby allowing to
significantly improve its ability of constraining cosmology
\citep[e.g.][]{Hu:1999,Huterer:2002,TakadaJain:2004}.

To address questions about the nature of dark energy and/or the
properties of gravity on cosmological scales, a number of ambitious
wide-field optical and infrared imaging surveys have been proposed:
the Panoramic Survey Telescope \& Rapid Response System
(Pan-STARRS\footnote{http://pan-starrs.ifa.hawaii.edu}), the Dark
Energy Survey (DES\footnote{http://www.darkenergysurvey.org}), the
Large Synoptic Sky Survey (LSST\footnote{http://www.lsst.org}), the
space-based Joint Dark Energy Mission (JDEM
\footnote{http://jdem.gsfc.nasa.gov}), and the EUCLID. However, there
are several sources of systematic errors inherent in weak lensing
measurements, and understanding the systematic errors is currently the
most important issue for achieving the full potential of planned
lensing surveys \citep[e.g.][]{Huterer:2010}.

One of the most dangerous systematic errors is the uncertainty in
estimating redshifts of source galaxies. Since it is practically
infeasible to obtain spectroscopic redshifts for the huge number of
imaging galaxies ($10^8$--$10^9$ galaxies for future surveys),
redshifts of galaxies need to be estimated from multi-band photometry
-- the so-called photometric redshifts (hereafter photo-$z$). Both
statistical errors and systematic biases in the relation between
photometric and spectroscopic redshifts need to be well controlled
(e.g. a sub-percent level for the bias for future surveys) in order
not to have any serious biases in cosmological parameters comparable
with the apparent statistical errors
\citep{Huterer.etal:2006,Ma.etal:2006}. Understanding the properties
of photo-$z$ errors is also important in exploring an optimal survey
design given the goal of achieving the desired level cosmological
constraints; depth vs number of filters vs area surveyed.  Given these
research backgrounds there are recent studies on photo-$z$ requirement
studies based on real data
\citep{Abdallaetal:2008,Limaetal:2008,Cunhaetal:2009}.

In this paper we would like to focus on an issue of how to identify
and remove catastrophic redshift errors -- the case that photometric
redshift is grossly misestimated \citep[also see][for the similar
study]{BernsteinHuterer:2009}. This can be done by monitoring the
posterior likelihood function of photo-$z$ estimation for each
galaxies. The important fact is that future surveys are planned to use
the sample variance limited regime in cosmic shear information rather
than the shot noise regime in order to constrain cosmology. Therefore
one can discard a large fraction of galaxies whose photo-$z$
estimations are less reliable \citep{Jainetal07}. Thus it would be
worth addressing how to construct a galaxy subsample suitable for
lensing experiments.  Having such a subsample of galaxies with
reliable photo-$z$ estimates may also relax requirements on a
spectroscopic training set to calibrate the residual photo-$z$
errors. In this paper we will address these issues by using the mock
catalog of photometric galaxies constructed based on the COSMOS
photo-$z$ catalog \citep{Ilbert.etal:2009} that provides the currently
most reliable photo-$z$ catalog calibrated with 30 bands data and
spectroscopic subsample.

This paper is organized as follow.  In Section~\ref{sec:cps}, we
briefly overview the theory of weak lensing.  In
Section~\ref{sec:data} we describe the details on how to make our
simulated mock catalog of photometric galaxies based on the COSMOS
data. In Section~\ref{sec:photoz} we use the simulated catalog to
assess the performance of photo-$z$ estimation assuming survey
parameters on depth and filter set, which are closely chosen to
resemble the Subaru Hyper Suprime-Cam(HSC) Survey. In
Section~\ref{sec:results} we show the main results: we use the
simulated photo-$z$ catalog to implement hypothetical weak lensing
experiment, paying particular attention to how to construct a galaxy
subsample which is defined such that it has minimal impact of the
photo-$z$ errors on cosmological
parameters. Section~\ref{sec:conclusion} is devoted to summary and
discussion. Unless explicitly stated we assume the concordance
$\Lambda$CDM model consistent with the WMAP 5-year results
\citep{Komatsu.etal:2009}.

%%%%%%%%%%%%%%%%%%%%%%%%%%%%%%%%%%%%%%%%%%%%%%%%%%%%%%%%%%%%%%%%%%%%%%%%%%%%%%%
\section{Preliminaries}
\label{sec:cps}
%%%%%%%%%%%%%%%%%%%%%%%%%%%%%%%%%%%%%%%%%%%%%%%%%%%%%%%%%%%%%%%%%%%%%%%%%%%%%%%

In this section we briefly review the basics of cosmic shear
tomography.  Throughout this paper we work in the context of a
spatially flat cold dark matter model for structure formation.

%------------------------------------------------------------------------------
\subsection{Convergence Power Spectrum}
%------------------------------------------------------------------------------
Gravitational shear can be simply related to the lensing convergence:
the weighted mass distribution integrated along the line of sight
\citep[e.g. see][for a thorough review and references
therein]{BartelmannSchneider01}.  Photometric redshift information on
source galaxies allows us to subdivide galaxies into redshift bins,
enabling more cosmological information to be extracted, which is
referred to as lensing tomography
\citep[e.g.][]{Hu:1999,Huterer:2002,TakadaBridle:2007}.  In the
context of cosmological gravitational lensing, assuming the flat-sky
approximation and the Limber's approximation, the lensing power
spectrum of the $i,j$-th tomographic bins can be expressed as
\begin{equation}
  P^{\kappa}_{ij}(\ell)
  =
  \int_{0}^{\infty} dz
  \frac{W_i(z)W_j(z)}{\chi^2(z)H(z)}
  P_\delta\!\left(k=\frac{l}{\chi}; z\right)
  \label{eq:convergence}
\end{equation}
where $H(z)$ is the Hubble expansion rate, $\chi$ is the comoving
angular diameter distance up to redshift $z$, and  
$P_\delta(k,z)$ is the three-dimensional matter power spectrum
at scale $k$ and at redshift $z$.  The lensing weight function
$W_{(i)}(\chi)$ in the $i$-th tomographic redshift bin, defined to lie
between the redshifts $z_i$ and $z_{i+1}$, is given by
\begin{equation}
  W_i(z)
  =
  \frac{3}{2}\Omega_{\rm m0} H^2_0 g_i(z)(1+z),
\label{eqn:weight}
\end{equation}
and 
\begin{eqnarray}
g_i(z)&=&
\left\{
\begin{array}{ll}
{\displaystyle
  \chi(z)
  \int_{{\rm max}(z,z_i)}^{z_{i+1}}\!\!dz'
\frac{n(z')}{\bar{n}_i}
\left[1-\frac{\chi(z)}{\chi(z')}\right]},
& 
z<z_{i+1}\nonumber\\
0,& 
z>z_{i+1}
\end{array}
\right.
\label{eqn:weight2}
\end{eqnarray}
where $n(z)$ is the redshift distribution of galaxies, and $\bar{n}_i$
is the average number density of galaxies residing in the $i$-th
tomographic bin (or the redshift range $z=[z_i,z_{i+1}]$):
$\bar{n}_i=\int_{z_i}^{z_{i+1}}dz' n(z')$.

In practice the power spectrum measured from a galaxy survey has shot
noise contamination arising from the finite sampling of galaxy
images. Hence the measured power spectrum becomes
\begin{equation}
C_{ij}^\kappa(l)=P^\kappa_{ij}(l)+\frac{\sigma_\epsilon^2}{\bar{n}_i}
\delta^K_{ij}, 
\end{equation}
where $\sigma_{\epsilon}$ is the rms intrinsic ellipticities per
component and $\delta_{ij}^K$ is the Kronecker delta symbols;
$\delta^K_{ij}=1$ when $i=j$, otherwise $\delta^K_{ij}=0$.  

Note that the distribution $n(z)$ appearing in Eq.~(\ref{eqn:weight})
denotes the underlying {\em true} redshift distribution of galaxies
used in lensing analysis.  However, the distribution needs to be
inferred from photo-$z$ information of individual galaxies available
from multi-color imaging data sets. This generally causes biases in
the lensing power spectrum in the presence of photo-$z$ errors. As
long as tomographic redshift bins are broad enough, $O(10^7)$ galaxies
are available in each bin for a Subaru-type survey with $\sim$1000
degree$^2$ sky coverage. Therefore the statistical errors of photo-$z$
are not problematic: the lensing power spectrum is primarily sensitive
to the mean redshift of source galaxies. Instead, a precise knowledge
of the mean redshift in each tomographic bin is required not to have
any significant biases in best-fit parameters compared to the
statistical errors, as studied in \cite{Huterer.etal:2006}.

To assess the required photo-$z$ accuracies for lensing tomography, an
important fact we should keep in mind is the lensing measurement for
planned wide-field surveys is not shot noise limited. Hence a
significant fraction of galaxies with ill-defined photo-$z$'s can be
discarded, without severely degrading parameter accuracies
\citep{Jainetal07}. With these considerations in mind we will in the
following address how to define an adequate subsample of galaxies for
a given multi-color data set.
\begin{figure}
  \plotone{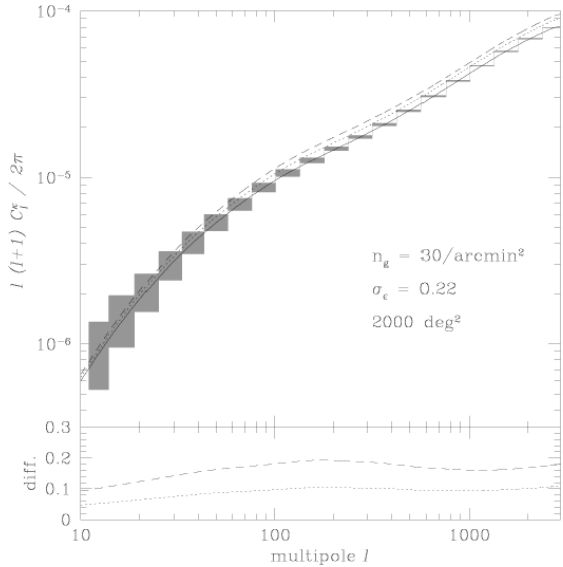}
  \caption[convergence power spectrum] { 
    The solid curve shows the lensing shear power spectrum assuming
    the concordance $\Lambda$CDM model and the galaxy redshift
    distribution with mean redshift $\langle z\rangle=1.13$ (see text
    for the details).  The dashed curve shows the resulting spectrum
    when the mean redshift is shifted by $5\%$, while the dotted curve
    shows the spectrum when the dark energy equation of state
    parameter is changed to $w=-1.2$.  The shaded boxes around the
    fiducial power spectrum show the expected $1\sigma$ error at each
    multipole bins assuming $\Omega_{\rm S}=2000$deg$^2$,
    $\bar{n}_g=30$ arcmin$^{-2}$, and $\sigma_\epsilon=0.22$ for
    survey area, the average number density of galaxies and the rms
    intrinsic ellipticities, respectively.  The bottom panel shows the
    relative differences of power spectra with respect to the fiducial
    spectrum.
    \label{fig:cl}
  }
\end{figure}
Figure~\ref{fig:cl} gives a quick summary of the impact of redshift
uncertainty on the lensing power spectrum for no tomography case,
i.e. a single redshift bin. Here for simplicity we assumed the
redshift distribution given by the analytical form, $n(z)\propto
z^2\exp[-(z/z_0)^2]$ with $z_0=1$, corresponding to the mean redshift
$\langle z\rangle = 2z_0/\sqrt{\pi}\simeq 1.12$. (Note that the
following results are all computed using simulated galaxy catalogs
that have different redshift distributions). The plot shows that a
$5\%$ change in the mean redshift causes a $10\%$-level change in the
power spectrum amplitude, and the amount of the change varies with
multipoles due to the projection of the nonlinear matter power
spectrum. This change can be compared with the effect of dark energy
equation of state and the statistical errors at each multipole bin
expected for the power spectrum measurement. Clearly such a bias in
the mean redshift is problematic for planned surveys.

%%%%%%%%%%%%%%%%%%%%%%%%%%%%%%%%%%%%%%%%%%%%%%%%%%%%%%%%%%%%%%%%%%%%%%%%%%%%%%%
\section{A Simulation of Photometric Galaxy Catalog}
\label{sec:data}
%%%%%%%%%%%%%%%%%%%%%%%%%%%%%%%%%%%%%%%%%%%%%%%%%%%%%%%%%%%%%%%%%%%%%%%%%%%%%%%
To assess the impact of photo-$z$ errors on cosmic shear tomography,
we take the following procedure. First we simulate a mock catalog of
galaxies that contain information on {\rm true} redshifts, magnitudes
in each filters and spectral energy distribution (SED) for survey
parameters we consider.  Then we estimate photometric redshifts
(hereafter often photo-$z$) for each simulated galaxies from its
colors, yielding the photo-$z$ catalogs.

A quick summary of the procedures used in making the mock photometric
catalog is as follows:
\begin{enumerate}
\item{} 
  Based on the results of COSMOS photo-$z$ catalog
  \citep{Ilbert.etal:2009}, we first model the redshift distribution
  of galaxies as a function of the $i$-band magnitudes down to
  $i=25.8$ (see \S~\ref{ssec:lf}).
\item{} 
  Use the synthetic galaxy spectral model, GISSEL98, to generate
  a set of SED templates for each type of galaxy, where the age and
  star formation history are randomly varied (see \S~\ref{ssec:spt}).
\item{} 
  Use the {\em HyperZ} code to generate a mock photometric catalog
  of galaxies in which the spectral energy distribution and
  redshift are assigned to each galaxy. In doing this the catalog is made
  by imposing the conditions 
  that the catalog satisfy the redshift-magnitude
  relation as well as reproduce an appropriate mixture of galaxy SED types
  which is consistent with the COSMOS galaxy population (see
  \S~\ref{ssec:hyperz}).
\item{} 
  For a given set of filters, compute apparent magnitudes in
  each filter for each simulated galaxy by taking into account the
  filter transmission curve and the redshifted spectrum (see
  \S~\ref{ssec:filter}). The statistical magnitude errors are also
  added to the magnitude of each filter (see \S~\ref{ssec:magerr}).
\end{enumerate}
To make a realistic simulated catalog, we assume survey parameters
(depth, filter transmission curves, and so on) that are chosen to well
resemble the planned HSC survey. We also consider the external imaging
data sets of $u$-band and/or NIR to study how combining the different
colors improves photo-$z$ accuracies.

In this paper we use the mock galaxy catalog containing about $10^5$
galaxies in the range $20 < i < 25.8$.

In the following subsections we will describe the details of each
procedure above, and a reader who is more interested in the results can
skip these subsections and go to \S~\ref{sec:photoz}.

%------------------------------------------------------------------------------
\subsection{Magnitude-Redshift Relation}
\label{ssec:lf}
%------------------------------------------------------------------------------
%
\begin{figure}
  \plotone{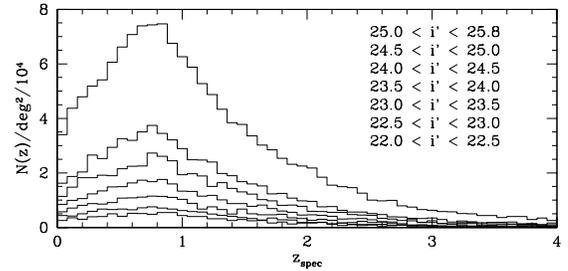}
  \caption{ The redshift distribution of our simulated galaxies
    (containing about $10^5$ galaxies) as a function of the $i$-band
    magnitude ranges as indicated by the labels. Note that the mock
    catalog is constructed so as to reproduce the redshift
    distribution found in the COSMOS galaxies with $i<25$. The
    galaxies with $25<i<25.8$ are simulated by extrapolating the
    COSMOS results down to the fainter magnitudes (see text for the
    details).
    \label{fig:cat}}
\end{figure}
\begin{figure}
  \plotone{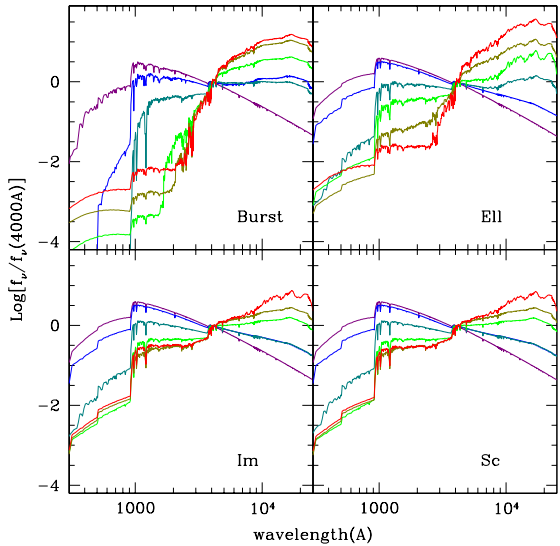}
  \caption{ Plotted are synthetic galaxy spectra,
    $f_{\nu}=\lambda^2f_{\lambda}$ normalized at $\lambda=4000$\AA.
    For each panel, from top to bottom at $10000$\AA, galaxy ages are
    10Gyr, 5Gyr, 1Gyr, 100Myr, 10Myr and 1Myr, respectively.  From the
    top-left to bottom-right panels, the galaxy SED types are
    star-burst (SB), elliptical (Ell), irregular (Im), and spiral
    (Sc), respectively.
    \label{fig:spt}}
\end{figure}

To make a mock catalog we need to properly take into account the
redshift distribution of galaxies, which varies with the range of
magnitudes considered.  For example, fainter galaxies are
preferentially at higher redshifts.  This is the so-called
magnitude-redshift relation.  We use the magnitude-redshift relation
estimated from the COSMOS catalog of galaxies selected by the Subaru
$i$-band magnitudes \citep{Ilbert.etal:2009}.  The COSMOS catalog
provides currently the most accurate photometric redshifts because the
photo-$z$ are estimated from 30 broad, intermediate, and narrow bands
covering from UV, optical to mid infrared. Also the photo-$z$
estimates are calibrated by the spectroscopic
sub-sample. \cite{Ilbert.etal:2009} studied subsamples of photo-$z$
galaxies for different limiting magnitudes and showed that the
resulting redshift distributions are well fitted by the polynomial
form:
\begin{equation}
  n(z)
  =
  A \frac{z^{a b}+z^{a}}{z^{b}+c}, 
  \label{eq:CFHTng}
\end{equation}
where $A$, $a, b $ and $c$ are the fitting parameters. The best-fit
parameters for different magnitudes in the range $i=[22,25]$ are given
in Table 2 in \cite{Ilbert.etal:2009}.  The COSMOS data is deep enough
in the $i$-band, and safely considered as a magnitude limit sample for
$i<25$. The COSMOS catalog also includes information on the angular
number counts of galaxies for a given magnitude range as well as on
the estimated galaxy SED type for each galaxy. To model a
hypothetically deeper survey we are interested in, we extrapolate the
fitting parameters to obtain the redshift distribution for fainter
galaxies down to $i=25.8$.

We thus generate a mock $i$-band photometric catalog of galaxies such
that the resulting catalog satisfies the magnitude and redshift
relation for different ranges of $i$-band
magnitudes. Figure~\ref{fig:cat} shows the magnitude-redshift
distributions for the simulated catalog containing about $10^5$
galaxies.

%------------------------------------------------------------------------------
\subsection{Synthetic Spectral Models}
\label{ssec:spt}
%------------------------------------------------------------------------------
For a given simulated galaxy labeled with some $i$-band magnitude and
redshift $z$, we need to model the spectral energy distribution from
which the apparent magnitudes can be computed for a given set of
filters. We use the publicly available library GISSEL98
\citep{BruzualCharlot:1993,BC2003,Bolzonella.etal:2000} to model the
synthetic galaxy spectrum.  The galaxy SEDs are generated to represent
from early- to late-type SEDs (elliptical, S0, Sa, Sb, Sc, Sd, Im and
starburst).  To model these populations -- composite stellar
populations (CSPs) -- the single stellar population (SSP) is convolved
with a model star formation history:
\begin{equation}
  f_{\rm CSP}(t)
  =
  \int_0^t
  \psi(t-\tau)f_{\rm SSP}(\tau) d\tau. 
\end{equation}
Note that the SSP is modeled in \cite{BruzualCharlot:1993}, with the
initial stellar mass function given in \cite{MillerScalo:1979}. The
function $\psi(t)$ is the star formation rate at galaxy age $t$. We
assumed $\psi(t) \propto \exp(-t/\tau)$ with $\tau=1, 2, 3, 5, 15, 30$
Gyr for elliptical, S0, Sa, Sb, Sc, and Sd galaxies, respectively.
For a starburst galaxy, the star formation is instantaneously
occurred, while $\psi={\rm constant}$ for an irregular (Im) galaxy.
The metalicity is self-consistently evolved with galaxy age, and we
checked that different models of metalicity little change the
photo-$z$ estimates \citep[also see][]{Bolzonella.etal:2000}. We
randomly chosen the age of each simulated galaxy from 221 different
ages in the range of $t=[0,20]$~Gyr, where the age interval is done
according to GISSEL98.

Figure~\ref{fig:spt} demonstrates simulated SEDs for starburst,
elliptical, irregular, and spiral galaxies for 6 different ages.  The
dust extinction is modeled following \cite{Calzetti.etal:2000} with
$A_V$ in the range [0,2.0].

%------------------------------------------------------------------------------
\subsection{Mock Galaxy Catalog}
\label{ssec:hyperz}
%------------------------------------------------------------------------------
To make a mock galaxy catalog containing various galaxy populations,
we used the publicly available code {\it HyperZ}\footnote{\sf
http://webast.ast.obs-mip.fr/hyperz/}. In doing this we need to
account for an appropriate mixture of different galaxy SED types. We
employed the composition (SB, E, S, Im)=(0.52, 0.035, 0.40, 0.045)
over all the redshift range, which is chosen so as to match the
composition of best-fit galaxy SED types found from the COSMOS catalog
with $i<25$ 
\footnote{\sf http://cosmos.astro.caltech.edu/data/index.html}.  For
this purpose the command {\it make\_catalog} in {\it HyperZ} was
slightly modified in such a way that the resulting catalog satisfies
the magnitude-redshift relations for each magnitude range in
\S~\ref{ssec:lf} and the assumed composition of galaxy SED types,
because the original {\it make\_catalog} generates a catalog that
redshift, reference magnitude, age, galaxy SED type and the amount of
dust extinction are randomly assigned to each galaxy.

%------------------------------------------------------------------------------
\subsection{Photometric Magnitudes}
\label{ssec:filter}
%------------------------------------------------------------------------------
Once the spectral energy distribution is specified for each simulated
galaxy at redshift $z$, it is straightforward to compute the apparent
magnitudes for a given set of filters taking into account the
redshifted spectrum at observed wavelengths. The photo-$z$ estimate is
sensitive to the details of observational parameters: the wavelength
coverage, the transmission curve of a given filter, the exposure time,
the limiting magnitude, and so on. We consider the parameters that
match those of the planned Subaru HSC survey: our default filter set
is $g'r'i'z'y'$ (hereafter the prime superscripts are sometimes
omitted), and the $5\sigma$ limiting magnitudes ($2^{\prime\prime}$
aperture) are set to $g=26.5$, $r=26.4$, $i=25.8$, $z=24.9$, and
$y=23.7$, respectively, assuming the exposure time of 15 minutes for
each pass-band, 3 days from new moon, and 1.2 airmass at the Subaru
Telescope site\footnote{The details can be found from {\sf
http://www.naoj.org/Observing/Instruments/SCam/index.html} and
\citep{Miyazakietal:2002}}.

We also study how adding other bands, $u$-band data and NIR data, into
the optical data above can improve photo-$z$ accuracies. Having a
wider wavelength coverage helps break degeneracies in photo-$z$
estimates, more exactly helps discriminate the Lyman break and 4000
angstrom break, from the multi-color data.  We here consider the
$u$-band data that can be delivered from CFHT, and also the
$J,H,K_s$(hereafter $K$) bands of planed VIKING (VISTA Kilo-Degree
Infrared Galaxy) survey.  The 5$\sigma$ limiting magnitudes are
$u=25.0$, $J=22.1$, $H=21.5$ and $K = 21.2$, respectively.
The set of filters and the depths we consider in this paper are
summarized in Table~\ref{tab:filters}.

%AJN
Detailed study for the optimal filter parameters, for example, the
number of filters, filter resolution, in terms of minimizing the
outlier fraction or photo-z scatters are found in
\cite{Jouvel.etal:2010}.

\begin{table}
  \begin{center}
  \caption{Filters and limiting magnitudes ($5\sigma$)
    \label{tab:filters}}
  \begin{tabular}{l | crrcc}\hline\hline
    Filter           & 
    Survey           & 
    $\lambda_c$(\AA) & 
    FWHM(\AA)        & 
    ABmag            & 
    T$_{\exp}$(sec) \\
    \hline
    $u*$    & CFHTLS & 3752 & 740     & 25.0 & 900 \\
    $g'$    & HSC    & 4814 & 1120    & 26.5 & 900 \\
    $r'$    & HSC    & 6279 & 1370    & 26.4 & 900 \\
    $i'$    & HSC    & 7687 & 1500    & 25.8 & 900 \\
    $z'$    & HSC    & 9143 & 1330    & 24.9 & 900 \\
    $y'$    & HSC    & 9923 & 490     & 23.7 & 900 \\
    $J$     & VIKING &12578 & 1713    & 22.1 & 420 \\
    $H$     & VIKING &16581 & 2828    & 21.5 & 420 \\
    $K$     & VIKING &21790 & 2828    & 21.2 & 420 \\
    \hline\hline
  \end{tabular}
  \end{center}
\end{table}
%

%------------------------------------------------------------------------------
\subsection{Magnitude Errors}
\label{ssec:magerr}
%------------------------------------------------------------------------------
Finally we include statistical errors in the apparent magnitudes.
Assuming the sky noise limit, we simply model this magnitude error as
Gaussian fluctuations with width given by
\begin{equation}
  \Delta m
  \simeq
 2.5\log \left( 1+\frac{1}{SN} \right),
  \label{eq:magerr}
\end{equation}
where $SN$ is the signal-to-noise ratio for a given galaxy; $SN$ is
computed once its apparent magnitude and the depth in the filter are
given. The magnitude error is computed as follows. First, the sky
noise is added to the observed flux of a galaxy, causing a deviation
from the true flux as $f^{\rm obs}=f_0+\Delta f=f_0(1+1/SN)$. Then the
magnitude error above is computed as $\Delta m=-2.5\log (1+1/SN)$
because $m_0+ \Delta m=-2.5\log f_0(1+1/SN)+{\rm constant}$. Exactly
speaking, even for a Gaussian sky noise, the magnitude error does not
obey a Gaussian distribution due to the log-mapping. However, the
Gaussian approximation on $\Delta m$ holds for galaxies with
sufficiently high $SN$ values, which we will assume for the following
results.

Note that a galaxy, which has its apparent magnitude near the limiting
magnitude, may be excluded from or included in the sample in the
presence of the magnitude error.  While our simulated galaxies are all
$i$-band selected, some galaxies may have apparent magnitudes below
the detection limit in other pass-bands. We use such an upper limit on
the flux in the photo-$z$ estimate, which improves the photo-$z$'s to
some extent.  In doing this the flux for such an undetected galaxy in
a given filter is set to the magnitude corresponding to the halved
limiting flux $f_{\rm lim}/2$.  Also we note that the systematic
offset of photometry may cause an additional uncertainty on magnitude
measurement, which in this paper is ignored.  Such a zero-point
magnitude offset can be, for example, calibrated by using a
spectroscopic redshift subsample \citep{Ilbert.etal:2009}.

According to the procedures described above we made a mock photometric
catalog containing about $10^5$ galaxies down to the magnitude
$i=25.8$.  Although we tried to make a realistic mock catalog based on
the COSMOS catalog, some of our treatments may be still
oversimplified: for example, we assumed a single stellar population
for galaxy SEDs.  The simplified assumptions may make our results
somewhat optimistic.  A more accurate way to overcome these obstacles
is using the real data including spectra for a representative
subsample of imaging galaxies. However, such a spectroscopic data
especially for faint galaxies of interest is still limited, and this
is our future work.

%%%%%%%%%%%%%%%%%%%%%%%%%%%%%%%%%%%%%%%%%%%%%%%%%%%%%%%%%%%%%%%%%%%%%%%%%%%%%%%
\section{Method: Photometric Redshift and Parameter Bias}
\label{sec:photoz}
%%%%%%%%%%%%%%%%%%%%%%%%%%%%%%%%%%%%%%%%%%%%%%%%%%%%%%%%%%%%%%%%%%%%%%%%%%%%%%%
Now we use the mock photometric catalog of $i$-band selected galaxies to
assess the performance of photo-$z$ estimates in the context of weak
lensing tomography experiment.

%------------------------------------------------------------------------------
\subsection{Photometric Redshift Estimation}
\label{ssec:photoz}
%------------------------------------------------------------------------------
By combining multi-passband magnitudes of a given imaging galaxy, its
redshift can be estimated without spectroscopic observation -- the
so-called photo-$z$.  There are various techniques that have been
developed: the template fitting method
\citep{Sawicki.etal:1997,Bolzonella.etal:2000}, the template method
combined with prior information (magnitude prior and so on)
\citep{Benitez:2000,Mobasher.etal:2004}, the method including a
self-calibration based on a training spectroscopic set \citep[][and
see references therein]{Collister.Lahav:2004}.

In this paper we use the publicly available code, {\it Le Phare} 
\footnote{{\sf http://www.cfht.hawaii.edu/\~{}arnouts/lephare.html}}, which
is a template fitting method. 
The photo-$z$ for each galaxy is estimated based on the $\chi^2$
fitting: 
\begin{equation}
  \chi^2
  =
  \sum_{i}^{N_{f}}
  \frac{
  \left[
    f_i^{\rm obs} - \alpha f(T,z,E)
    \right]^2
  }%
  {\sigma_i^2},
\label{eq:photoz}
\end{equation}
where $f_i^{\rm obs}$ is the observed flux in the $i$-th filter,
$f(T,z,E)$ is the model flux which is given as a function of galaxy
SED type ($T$), redshift ($z$) and the amount of dust extinction
($E$), and $\sigma_i$ is the magnitude error. Note that galaxy SED
type is modeled according to the method described in
\S~\ref{ssec:spt}. The summation runs over the number of filters
considered ($N_f$).  The extra factor parameter $\alpha$, which is the
same in all the filters, is introduced in Eq.~(\ref{eq:photoz})
because the photo-$z$ is estimated only from colors, the relative
amplitudes of fluxes in different filters, not from the absolute
fluxes.  Therefore there are ($N_f$-1) constraints given the data of
$N_f$ filters.  The best-fit redshift parameter, i.e. the best-fit
photo-$z$, is obtained by minimizing the $\chi^2$ value with varying
other model parameters.

If the location of spectral features such as Lyman break and 4000\AA~
break is captured given the wavelength coverage of observed filters
and the magnitude depths taken, the redshift is robustly estimated. On
the other hand, a misidentification of the spectral features causes a
degeneracy in redshift estimation, often yielding multiple solutions
at low and high redshifts.  Hence the photo-$z$ method based on broad
band photometry generally yields a large fraction of outliers, where
the best-fit photo-$z$ can be far from the true redshift.  To quantify
the photo-$z$ accuracy for each galaxy we will use the following two
quantities: (1) the goodness-of-fit parameter for the template
fitting, and (2) the width of likelihood function of redshift
estimation.

The goodness-of-fit for the template fitting of a given galaxy may be
defined as
\begin{equation}
  \chi^2_\nu
  \equiv 
  \frac{\chi^2_{\rm min}(z_{\rm bf})}{N_{f}-1},
  \label{eq:redchi2}
\end{equation}
where $\chi^2_{\rm min}$ is the minimum $\chi^2$ value for the
best-fit model and redshift, $z_{\rm bf}$ is the best-fit redshift and
$N_f-1$ is the number of colors available. Note that the quantity
above, $\chi^2_\nu$, is not the same as the reduced $\chi^2$, which is
defined as the number of constraints minus the number of model
parameters. The number of model parameters are equal for all the
galaxies, so $\chi^2_\nu$ gives a measure of the goodness-of-fit.

\begin{figure}[htbp]
  \plotone{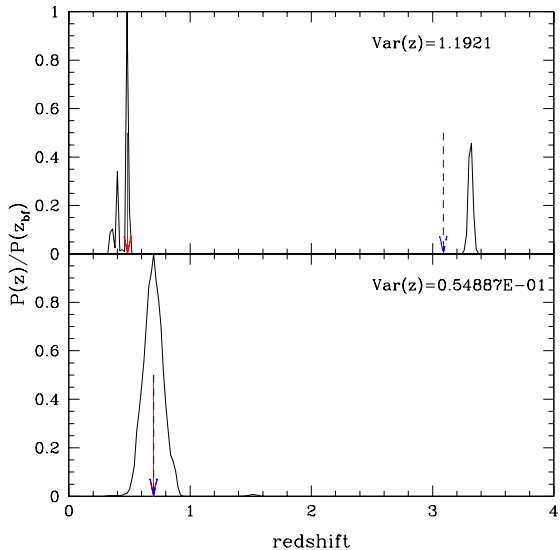}
  \caption{ 
    Examples of the photo-$z$ fitting for two simulated
    galaxies. The upper panel shows an example of the photo-$z$
    outlier, i.e. ill-defined photo-$z$ estimate, while the lower
    panel shows an example of the reliable photo-$z$ estimate. In each
    panel, the solid curve shows the likelihood function of redshift
    parameter, $p(z)\propto \exp[-\chi^2/2]$, and the solid (red) and
    dashed (blue) arrows denote the best-fit redshift and the true
    redshift, respectively. The width of the likelihood is quantified
    by ${\rm Var}(z)$, defined by Eq.~(\ref{eq:secmom}), and the value
    for each simulated galaxy is denoted in the upper-right corner of
 each panel.
  \label{fig:pdf}}
\end{figure}

We also use the width of likelihood function of redshift estimation
for each galaxy defined as
\begin{equation}
  {\rm Var}(z)
  \equiv
  \int_0^\infty\!dz~ (z - z_{\rm bf})^2 
  p(z)
  \frac{1}{(1+z_{\rm bf})}, 
  \label{eq:secmom}
\end{equation}
where $p(z)$ is the likelihood function given as $p(z)\propto
\exp[-\chi^2(z)/2] $, which is normalized so as to satisfy
$\int\!dz~p(z)=1$.  We compute the likelihood function $p(z)$ by
fixing other model parameters to their best-fit values.  The
normalization factor $(1+z_{\rm bf})$ is introduced based on the fact
that the photo-$z$ accuracy scales with $(1+z)$.  Compared to
$\sigma(z)$, the local standard deviation of photo-$z$ estimation,
which is obtained from $\Delta \chi^2\le 1$, the quantity ${\rm
Var}(z)$ is sensitive to the outlier probability with $|z-z_{\rm
bf}|\gg 1$ due to the weight $(z-z_{\rm bf})^2$. Hence for galaxies
whose likelihood function has multiple peaks, i.e. multiple redshift
solutions, the quantity ${\rm Var}(z)$ tends to be larger.  The
similar quantities to ${\rm Var}(z)$ are also used in the previous
works \citep{Mobasher.etal:2004,Wolf:2009}, where the primary purpose
is to improve the photo-$z$ performance for a majority of galaxies. In
this paper we use the figure-of-merit quantity ${\rm Var}(z)$ mainly
for identifying photo-$z$ outliers.

Note that we will in the following results use $z_{\rm bf}$ to construct the
redshift distribution of galaxies. An alternative method, which may be
less sensitive to photo-$z$ outliers, is summing the photo-$z$
posterior likelihood function $p(z)$ over all the sampled galaxies to
obtain the overall redshift distribution
\citep[][]{Wittman:2009,Cunhaetal:2009}. Calibrating the mean redshift
distribution is another important issue to be carefully studied
\citep{MaBernstein:2008,Bordoloi.etal:2009},
but is beyond the scope of this paper. 

Figure~\ref{fig:pdf} demonstrates examples of the photo-$z$ fitting
for two simulated galaxies. It can be found that a galaxy which has
the ill-defined photo-$z$ estimate, i.e. the wider likelihood function
of redshift estimation, tends to have a larger value of ${\rm
Var}(z)$. In particular, even if the likelihood function locally has a
narrow peak around the best-fit redshift, therefore even if the
photo-$z$ error looks apparently small, the value ${\rm Var}(z)$
becomes larger if the likelihood has multiple peaks (i.e. the case of
multiple redshift solutions), as demonstrated in the upper panel.  On
the other hand a galaxy with reliable photo-$z$ estimate has a small
value of ${\rm Var}(z)$.

\begin{figure}
  \plotone{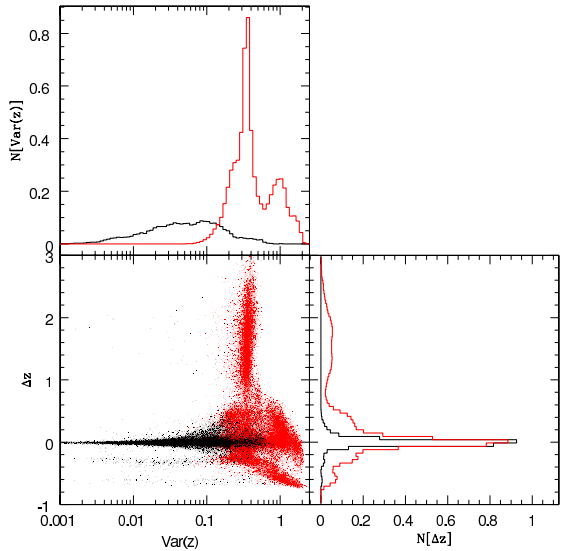} 
\caption{ The probability distributions of simulated galaxies in
  parameter space of ${\rm Var}(z)$ and $\Delta z$, where ${\rm Var}(z)$
  is defined in terms of the photo-$z$ likelihood by
  Eq.~(\ref{eq:secmom}) and $\Delta z$ denote the bias between photometric
  and true redshifts, $\Delta z\equiv (z_{\rm bf}-z_{\rm
  true})/(1+z_{\rm true})$. In each panel, black-thick dots or lines
  show galaxies whose photo-$z$ likelihood function has a single peak,
  while gray-thin dots or lines show galaxies that multiple peaks in
  their likelihood. It is clear that galaxies, which have greater
  photo-$z$ biases and have multiple photo-$z$ solutions in their
  likelihood, tend to have greater values of ${\rm Var}(z)$.
\label{fig:pdfscatter}}
\end{figure}

While the quantity ${\rm Var}(z)$ is rather empirically defined as an
indicator of photo-$z$ outliers, Figure~\ref{fig:pdfscatter} gives a
quantitative study of how ${\rm Var}(z)$ can characterize the photo-$z$
likelihood function. The figure shows the distributions of simulated
galaxies in the ${\rm Var}(z)$--$\Delta z$ plane, where $\Delta z$ is
the difference between true and photometric redshifts defined as $\Delta
z\equiv (z_{\rm bf}-z_{\rm true})/(1+z_{\rm true})$ for each
galaxy. According to the properties of their photo-$z$ likelihood
functions, we divide galaxies into two subsamples: one (black-thick
line) is defined from galaxies (about 25\% fraction of all the galaxies)
that have a single peak therefore a reliable photo-$z$ estimate, while
the other (gray-thin line) is from galaxies of multiple peaks,
respectively. Note that the second and higher-order peaks are defined
from local maxima of the likelihood that have heights higher than 10\% of the
first peak height. One can find from the figure that the quantity ${\rm
Var}(z)$ nicely separate galaxies that tend to have have greater
photo-$z$ biases and multiple solutions of photo-$z$'s, i.e. degenerate
photo-$z$ estimate; most of galaxies having ${\rm Var(z)}\simgt 0.1$
have multiple peaks.

%------------------------------------------------------------------------------
\subsection{Fisher matrix formalism}
\label{ssec:fisher}
%------------------------------------------------------------------------------
We will use the Fisher matrix formalism to estimate accuracies of
estimating parameters given the lensing power spectrum.  The Fisher
matrix is given by
\begin{equation}
F_{\alpha\beta}=\sum_{\ell=\ell_{\rm min}}^{\ell_{\rm max}}
\sum_{i,j,m,n}
\frac{\partial C^{\kappa}_{ij}(\ell)}{\partial p_\alpha}
{\bf C}^{-1}\!\!\left[
C^\kappa_{ij}(\ell),C^\kappa_{mn}(\ell)
\right]
\frac{\partial C^{\kappa}_{mn}(\ell)}{\partial p_\beta},
\label{eq:fisher}
\end{equation}
where $p_\alpha$ denotes a set of cosmological parameters, the matrix
${\bf C}$ denotes the covariance matrix, and ${\bf C}^{-1}$ denotes
the inverse matrix.  In this paper we simply use the covariance matrix
given by the first term in Eq.~(9) in \cite{TakadaJain09} assuming the
Gaussian errors on power spectrum measurements. The Gaussian error
assumption is adequate for our purpose because the impact of
non-Gaussian errors on parameter estimation is not significant as long
as a multi-parameter fitting is considered as shown in
\cite{TakadaJain09}. The marginalized $1\sigma$ error on the
$\alpha$-th parameter $p_\alpha$ is given by $\sigma^2(p_\alpha)=({\bf
F}^{-1})_{\alpha\alpha}$, where ${\bf F}^{-1}$ is the inverse of the
Fisher matrix.  Throughout this paper we set $l_{\rm min}=5$ and
$l_{\rm max}=3000$ as for the minimum and maximum multipoles in the
summation above.  Note that all the parameter forecasts shown below
are for the lensing tomography combined with the expected Planck
information, which is obtained simply by adding the two Fisher
matrices of lensing and CMB: ${\boldsymbol F}_{\rm
WL+CMB}={\boldsymbol F}_{\rm WL}+{\boldsymbol F}_{\rm CMB}$.

As explained around Eq.~(\ref{eq:convergence}), the lensing power
spectrum is sensitive to the underlying true redshift distribution of
galaxies, $n(z)$. For a multi-color imaging survey, however, the
distribution $n(z)$ needs to be estimated from the available photo-$z$
information.  In this procedure the photo-$z$ errors affect weak
lensing experiments. Most dangerous effect is a systematic bias in
parameter estimations: if the inferred redshift distribution has a
bias in the mean redshift compared to the true one, the redshift bias
may cause significant biases in cosmological parameters.
In order to quantify the biases in cosmological parameters caused by
photo-$z$ errors, we use the following method based on the Fisher
matrix formalism in \cite{HutererTakada:2005} \citep[also see
Appendix B of][for the detailed derivation]{Joachimi.Schneider:2009}:
\begin{eqnarray}
  \delta p_\alpha 
  &=&
  \sum_{\beta} 
      [{\boldsymbol F}_{\rm WL+CMB}^{-1}]_{\alpha\beta}
      \sum_{\ell}\sum_{i,j,m,n}
      \frac{\partial C^\kappa_{ij}(\ell)}{\partial p_\beta}
      \nonumber\\
  & &
      \times {\bf C}^{-1}[C^{\kappa}_{ij}(\ell),C^\kappa_{mn}(\ell')] 
      \left[
	C^{\kappa}_{mn}(\ell')-C^{\kappa,{\rm photo-z}}_{mn}(\ell')
	\right],
      \nonumber\\
  \label{eq:bias_theta_red}
\end{eqnarray}
where ${\boldsymbol F}^{-1}_{\rm WL+CMB}$ is the inverse of the Fisher
matrix, and $\delta p_\alpha$ denotes a bias in the $\alpha$-th
parameter, the difference between the best-fit and true values. In the
equation above, the spectrum $C_{mn}^\kappa(l)$ is the underlying true
power spectrum, while $C^{\kappa, {\rm photo-z}}_{mn}(l)$ is the
spectrum estimated from the redshift distribution inferred based on
the photo-$z$ information.  In the presence of the photo-$z$ errors,
generally $C^{\kappa}_{ij}\ne C^{\kappa,{\rm photo-z}}_{ij}$, thereby
causing a bias in parameter estimation.  We can compute both spectra,
$C^{\kappa}$ and $C^{\kappa,{\rm photo-z}}$ from a simulated galaxy
catalog for a hypothetical lensing survey.  Note that in
Eq.~(\ref{eq:bias_theta_red}), for simplicity, we have not considered
any other nuisance parameters that model other systematic effects such
as the shape measurement errors \citep[e.g.][]{Huterer.etal:2006} and
the inability to make precise model predictions arising from nonlinear
clustering and baryonic physics
\citep[][]{HutererTakada:2005,Rudd.etal:2008,Zentner.etal:2008}.

To compute the parameter forecasts we need to specify a fiducial
cosmological model and survey parameters.  Our fiducial cosmological
model is based on the WMAP 5-year results \citep{Komatsu.etal:2009}:
the density parameters for dark energy, CDM and baryon are
$\Omega_{\rm de}(=0.74)$, $\Omega_{\rm cdm}h^2(=0.1078)$, and
$\Omega_{\rm b}h^2(=0.0196)$ (note that we assume a flat universe);
the primordial power spectrum parameters are the spectral tilt,
$n_s(=1)$, and the normalization parameter of primordial curvature
perturbations, $A_s\equiv \delta_\zeta^2(=2.3\times 10^{-9})$ (the
values in the parentheses denote the fiducial model); the dark energy
equation of state parameter $w_0(=-1)$. We used the publicly available
code CAMB developed in \cite{camb} \citep[also see][]{cmbfast} to
compute the transfer function, and use the fitting formula in
\cite{Smithetal:2003} to compute the nonlinear mass power spectrum
from which the lensing power spectrum is computed over the relevant
range of angular scales.

Our fiducial survey roughly resembles the planned Subaru Hyper-Suprime
Cam Survey \citep{Miyazaki:2006}. We adopt the set of filters
($grizy$) and the depths in each filter as given in
\S~\ref{sec:data}. We will also study how the results change when the
hypothetical Subaru survey is combined with other surveys that deliver
the $u$-band data or/and the NIR data, which especially help improve
the photo-$z$ accuracies. The survey area is throughout assumed to be
$\Omega_{\rm s}=2000~$deg$^2$.  The redshift distribution of galaxies
is computed for an assumed subsample of galaxies based on the photo-$z$
information. 

%------------------------------------------------------------------------------
\subsection{Object selection and clipping of photo-$z$ outliers}
\label{ssec:objectselection}
%------------------------------------------------------------------------------
We may be able to construct a suitable subsample of galaxies in a
sense that the impact of photo-$z$ errors are minimized in order not
to have more than 100\% biases in cosmological parameters compared to
the statistical errors.  Hence a selection of adequate galaxies is
important for weak lensing: for example, this may be attained by
discarding galaxies with ill-defined photo-$z$'s.  However, the
important fact we should keep in mind is that, with discarding more
galaxies, the statistical accuracy of parameter estimation is degraded
due to the increased shot noise contamination in the power spectrum
measurement. Thus there would be a trade-off point in defining a
suitable galaxy subsample in terms of the parameter bias versus the
statistical error for a given survey.

We throughout this paper work on $i$-band selected galaxies assuming
that the $i$-band data is used for the lensing shape measurement as
often done in the previous lensing works.  For our simulated galaxies,
given the limiting magnitude $i=25.8$ at $5\sigma$ significance, a
sufficiently number of photometric galaxies are available: the number
density for total galaxies is 80 per square arcminutes.  However, all
the galaxies are not usable of lensing measurements. First, in order
to obtain a reliable shape measurement, galaxies used in the lensing
analysis need to be well resolved, requiring the galaxies to have
sufficient signal-to-noise ratios (say more than
$10\sigma$). Secondly, galaxies with ill-defined photo-$z$ estimates
are not useful because including such galaxies may cause a significant
bias in parameter estimation.

With the considerations above in mind, we will consider the following
object selections or their combinations to make parameter forecasts. 
\begin{itemize}
\item{}
  The restrictive range of $i$-band magnitudes: $22.5\le i\le
  25$. The range is a typical one used in the weak lensing analysis
  \citep[e.g.][]{Okabeetal:2009}. The faint-end magnitude cut may
  be imposed such that the selected galaxies have sufficiently high
  signal-to-noise ratios: $i=25$ corresponds to $S/N\simeq 10$ in
  our simulations. The bright-end magnitude cut is not important,
  but usually imposed in practice to avoid galaxies with saturated
  pixels.
\item{}
  The photo-$z$ selection. We select only galaxies that have 
  reasonably good photo-$z$ estimates by imposing a threshold on the
  goodness-of-fit of photo-$z$ estimation,
  $\chi^2_\nu\le 2$ (see Eq.~[\ref{eq:redchi2}]). The clipping threshold
  $\chi^2_\nu=2$ is not a unique choice.
  Rather we selected the value, as one
  working example.
\item{}
  The restrictive range of photo-$z$'s: $0.2\le z_{\rm bf}\le 1.5$.
  The spectral features of galaxies in this range of redshifts can be
  relatively well captured by the wavelength coverage of optical
  filters.  The lower redshift cut is introduced, because there is a
  strong degeneracy between galaxies at such low redshifts $z\simlt
  0.2$ and those at higher redshifts, especially in a case that the
  $u$-band data is not available or shallower than optical data as
  considered in this paper.
\item{}
  Clipping of photo-$z$ outliers. By discarding galaxies with
  ${\rm Var}(z)$ (see Eq.~[\ref{eq:secmom}]) greater
  than a given threshold, which turn out to be mostly photo-$z$
  outliers, we define a subsample from the remaining galaxies that have
  relatively reliable photo-$z$'s. In the following we study the
  performance of this clipping method by varying the clipping threshold
     values of 
  ${\rm Var}(z)$.
\end{itemize}
\begin{table}
  \begin{center}
    \caption{Fractions of galaxies included in each subsample
      \label{tab:dataset}}
    \begin{tabular}{l | rcc}\hline\hline
      Filters             & 
      $\chi^2_\nu<2$      & 
      $\cap 0.2<z_p<1.5$  & 
      $\cap 22.5<i'<25.0$ \\  \hline
      $grizy$     & 0.94  & 0.61          &  0.50    \\
      $ugrizy$    & 0.94  & 0.61          &  0.50    \\
      $grizyJHK$  & 0.89  & 0.57          &  0.47    \\
      $ugrizyJHK$ & 0.89  & 0.58          &  0.48    \\    \hline\hline
    \end{tabular}
    \tablecomments{ 
      The subsamples denoted as ``$\chi^2_\nu\le 2$''
      show included galaxies selected with $\chi^2_\nu\le 2$ in the
      photo-$z$ fitting for each combination of filters (see text for
      the details). The second and third columns show the results when
      further imposing the conditions on the ranges of photo-$z$'s and
      $i$-band magnitudes as denoted.}
  \end{center}
\end{table}

Table~\ref{tab:dataset} gives the fraction of remaining galaxies
compared to the original sample, where galaxies in each subsample are
selected with object selection criteria described above for a given
set of filters.  The column denoted by ``$\chi^2_\nu\le 2$'' shows the
fraction of galaxies selected when imposing the threshold on the
goodness-of-fit for each galaxy. It is clear that this clipping
discards only a small fraction of galaxies for all the cases of filter
combinations.  The second and third columns show the fractions when
further imposing the restricted ranges of photometric redshifts and
magnitudes, respectively.

%%%%%%%%%%%%%%%%%%%%%%%%%%%%%%%%%%%%%%%%%%%%%%%%%%%%%%%%%%%%%%%%%%%%%%%%%%%%%%%
\section{Results}
\label{sec:results}
%%%%%%%%%%%%%%%%%%%%%%%%%%%%%%%%%%%%%%%%%%%%%%%%%%%%%%%%%%%%%%%%%%%%%%%%%%%%%%%
In this section we show the main results of this paper using mock galaxy
catalogs. 

%------------------------------------------------------------------------------
\subsection{Photo-z Accuracy}
\label{ssec:photozacc}
\begin{figure}[htbp]
  \plotone{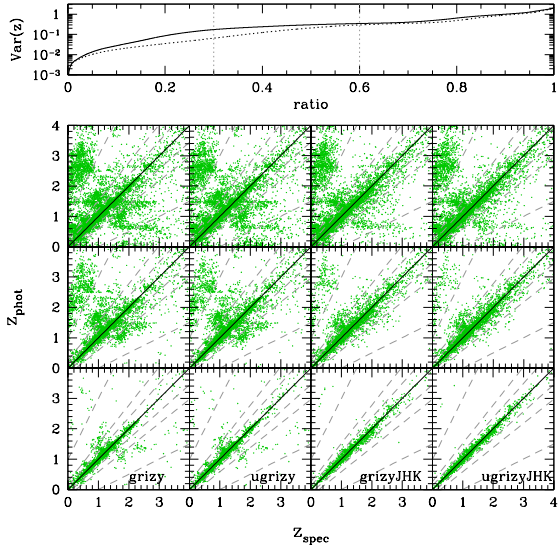}
  \caption{ 
    {\it Lower plot}: The scatter plots between photometric
    and true redshifts for our simulated galaxies. The different
    columns correspond to the results for different sets of filters as
    indicated in the bottom panels, while the different rows
    correspond to different object selections (see
    \S~\ref{ssec:objectselection} for the details). The upper-row
    panels show the results of samples containing all the galaxies
    with $i<25.8$ that have their photo-$z$ fits quantified as
    $\chi^2_\nu\le 2$ (see Eq.~[\ref{eq:redchi2}]). The middle- and
    bottom-row panels show the results of subsamples obtained by
    discarding 40\% and 70\% galaxies with ill-defined photo-$z$
    estimates, respectively. This is done by choosing the threshold
    value ${\rm Var}(z)$ (see Eq.~[\ref{eq:secmom}]) for each galaxy
    such that the desired fractions of galaxies are remained in the
    resulting subsamples. Note that, for illustrative purpose, only
    5\% representative galaxies in each subsample are shown in each
    plot.  {\it Upper panel}: The ratio of remaining galaxies that
    have ${\rm Var}(z)$ values greater than a given threshold denoted
    on the vertical axis. The solid curve shows the result for the set
    of filters, $grizy$, while the dotted curve shows the result when
    the NIR data $JHK$ are added (see Table~\ref{tab:filters} for the
    details). Note that the results are almost unchanged by further
    adding the $u$-band data. The vertical thin dashed lines denote
    the selections used in the lower plot, the criteria discarding
    40\% and 70\% galaxies with poor photo-$z$'s.
    \label{fig:scatter}}
\end{figure}

%------------------------------------------------------------------------------
Figure \ref{fig:scatter} shows the results using different galaxy
catalogs with various combinations of
the measured passbands (see Table~\ref{tab:filters}), 
$grizy$, $ugrizy$, $grizy$+$JHK$ and $ugrizy$+$JHK$ from the left- to
right-column panels, respectively.  Each panel in the top row shows
the photo-$z$ performance for the simulated objects with $i'<25.8$
($>5\sigma$), but selected by imposing the condition on the
goodness-of-fit $\chi^2_\nu<2$ (see Eq.~\ref{eq:redchi2}). It is clear
that, with broadband photometry alone, the photo-$z$ accuracy is
limited: a significant contamination of outliers is unavoidable, even
with including NIR- and/or $u$-band data.

The middle- and lower-row panels show the results obtained by further
discarding photo-$z$ outliers based on the clipping method with a
given threshold on the quantity ${\rm Var}(z)$
(Eq.~[\ref{eq:secmom}]). The thresholds are chosen such that 40\% or
70\% of the objects in each top-row panels, which tend to have
ill-defined photo-$z$'s, are discarded, respectively. It can be found
that the clipping method based on the photo-$z$ likelihood function of
each galaxy can efficiently discard galaxies with ill-defined
photo-$z$'s, especially when combined with the NIR and $u$-band data.

To be more explicit, the upper panel shows how much fraction of
galaxies are included in the subsample when discarding galaxies with
${\rm Var}(z)$ greater than a given threshold denoted on the vertical
axis.  The solid and dotted curves show the results without and with
the $JHK$ data in addition to the optical data, $grizy$. Note that
including the CFHT-type $u$-band data of the assumed depth, as given
in Table~\ref{tab:filters}, little changes the two results. For a
given particular value of ${\rm Var}(z)$, the dotted curve has more
remaining galaxies than the solid curve, implying that these NIR data
help improve the photo-$z$ accuracies on individual galaxy basis,
i.e. indicating that the shape of photo-$z$ likelihood function
becomes narrowed by adding the NIR data for most of galaxies.

\begin{figure}[htbp]
  \plotone{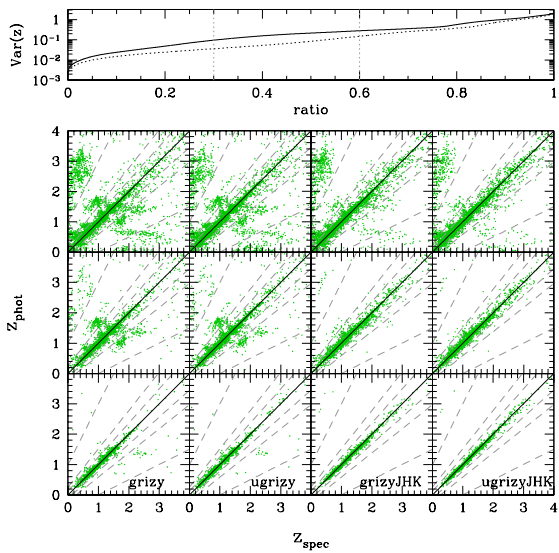} 
  \caption{ Similar to Figure~\ref{fig:scatter}, but
    for the different range of $i$-band magnitudes for object selection,
    $22.5<i<25$.
    \label{fig:scatter2}
  }
\end{figure}

Figure~\ref{fig:scatter2} shows the similar result to the previous
figure, but for brighter samples of galaxies, selected with $i<25$ or
equivalently with $S/N$ values greater than $10\sigma$ in
$i$-band. These brighter galaxies are more suitable for the accurate
shape measurement as discussed in \S~\ref{ssec:objectselection}. One
can find that the photo-$z$ accuracy is improved compared to
Figure~\ref{fig:scatter}.

%------------------------------------------------------------------------------
\subsection{Simulating Lensing Tomography: The Impact of Photo-$z$ Errors}
%------------------------------------------------------------------------------
%
\begin{figure*}[htbp]
 \begin{center}
  \epsscale{.8} 
  \plotone{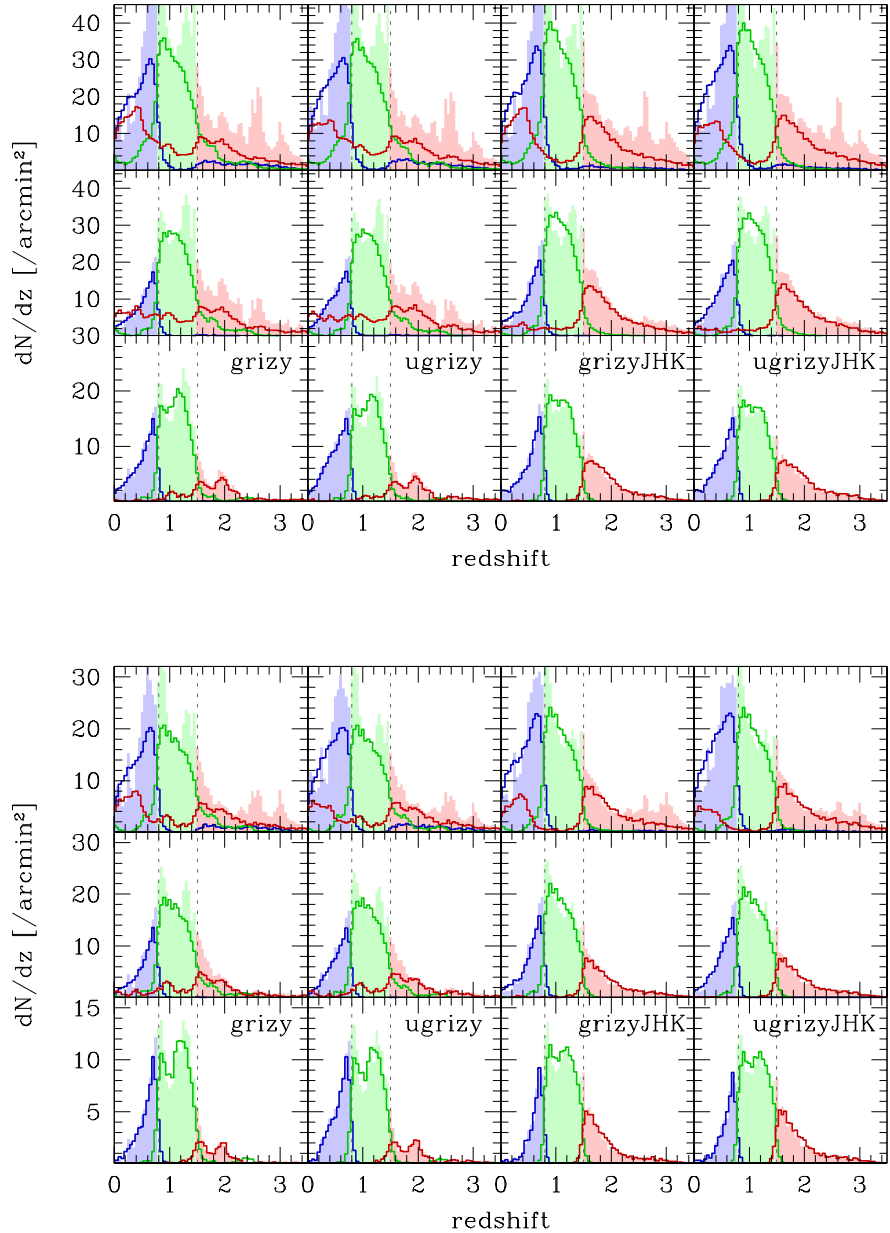} 
  \caption{ 
    The shaded histograms are the redshift distributions of
    galaxies based on their photo-$z$ information, i.e. the sharp cuts
    imposed on the photo-$z$'s: $0<z_p<0.8$, $0.8<z_p<1.5$ and
    $z_p>1.5$, indicated on the vertical dashed lines. In this case
    the horizontal axis denotes photo-$z$ values.  The solid-line
    histograms are the underlying true distributions, therefore the
    horizontal axis denotes true redshifts in this case.  The upper
    and lower panels are the results using the galaxy catalogs in
    Figures~\ref{fig:scatter} and \ref{fig:scatter2},
    respectively. The different panels are for different galaxy
    selections as in Figures~\ref{fig:scatter} and \ref{fig:scatter2}.
    \label{fig:nzsp}
  }
 \end{center}
\end{figure*}

We are now in a position to use the photo-$z$ galaxy catalogs,
constructed up to the preceding section, to make the trade-off
analysis on the number of galaxies within a sample versus how
``clean'' the tomographic redshift intervals are. Then we study the
impact of photo-$z$ errors on parameter estimation assuming the
hypothetical lensing tomography experiment.

Figure~\ref{fig:nzsp} shows the redshift distributions of each
tomographic bin made by subdividing the photo-$z$ galaxies into 3
intervals of photometric redshifts, $z_p<0.8$, and $0.8<z_p<1.5$ and
$z_p>1.5$. The redshift intervals are chosen such that each redshift
intervals contain similar number densities for the original mock
catalog of galaxies.  Note that the redshift binning is fixed in the
following analysis for simplicity, which helps to compare the results
of different galaxy catalogs. Also note that three redshift bins are
a minimal choice of lensing tomography for constraining the dark
energy equation of state parameter ``$w$'' to a reasonable accuracy by
efficiently breaking parameter degeneracies in the lensing power
spectrum \citep[e.g.][]{TakadaJain:2004}.

The upper plot in Figure~\ref{fig:nzsp} shows the tomographic redshift
distributions constructed from different photo-$z$ catalogs in
Figure~\ref{fig:scatter}, where different panels correspond to the
different sets of filters and the different clipping thresholds.  The
shaded regions in each panel show the photometric redshift
distributions of galaxies, which have sharp cutoffs in the
distributions due to the sharp redshift binning, while the solid line
histograms show the underlying true redshift distributions.  As can be
found from the top-row panels, if all the galaxies are used, the
resulting redshift distributions have significant overlaps between
different redshift bins due to a significant contamination of
photo-$z$ outliers for any combinations of filters.  On the other
hand, the middle- and bottom-row panels show that, when 40\% or 70\%
of photo-$z$ outliers are discarded by imposing the corresponding
thresholds on ${\rm Var}(z)$, respectively, the overlaps can be
increasingly reduced. In particular, when the optical data is combined
with the NIR data such that those expected from the VIKING survey, the
resulting subsamples have almost no overlap, if about $ 70\%$ of
galaxies are discarded.  We should note that such a clean redshift
binning can greatly reduce a possible contamination of the intrinsic
ellipticity alignments arising from the physically close pairs of
galaxies in the similar redshifts \citep{TakadaWhite:2004}.

The lower plot shows the similar results for higher $S/N$ samples with
$22.5<i<25$ whose scatter plots are seen in
Figure~\ref{fig:scatter2}. Again, if using the clipping method and
having a wider coverage of wavelengths, a clean subsample with almost
no overlap between redshift bins can be obtained.

\begin{figure}[htbp]
 \plotone{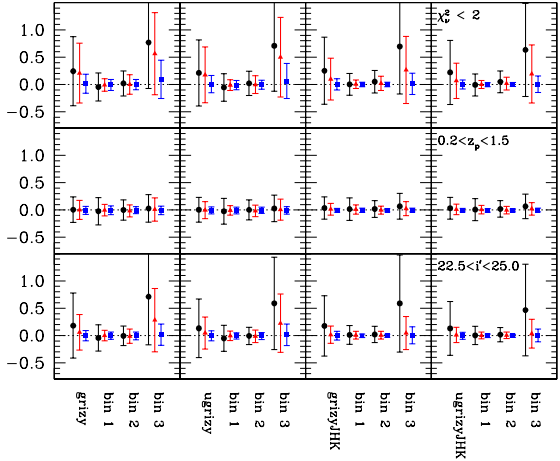}
 \caption{ 
   The bias between photometric and true redshifts, $\langle
   \Delta z \rangle$, in each tomographic redshift interval. The error
   around each point is the statistical error of the mean redshift,
   $\sigma(\langle \Delta z \rangle)$, in each redshift bin.  For
   illustrative purpose, the errors are for a survey area of 1
   sq. arcminutes, and the errors are smaller by a factor
   $\sqrt{2000\times60^2}\simeq 2700$ for our fiducial survey area of
   2000 sq. degrees.  The panels in different columns show the results
   for different sets of filters as indicated on the horizontal axis.
   The upper-row panels are for the subsamples where galaxies with
   $i<25.8$ are selected only with the condition $\chi^2_\nu<2$. The
   middle- and bottom-row panels are for the subsamples where the
   condition $0.2<z_p<1.5$ or $22.5<i<25$ is further imposed for the
   selection, respectively.  The round symbols in each panel show the
   results for the whole galaxy sample, while the triangle and square
   symbols are the results for the subsamples discarding 40\% and 70\%
   of galaxies with ill-defined photo-$z$'s, respectively, based on
   our clipping method.  In each panel the four symbols are for
   different redshift intervals: ``bin1'', ``bin2'', and ``bin3''
   correspond to the lowest, medium and highest redshift bins in
   Figure~\ref{fig:nzsp}, and the leftmost symbols are for the case of
   no tomography, i.e. a single redshift interval.
   \label{fig:bias_scatter}
 }
\end{figure}

As have been stressed several times, the weak lensing power spectra
are, to the zero-th order approximation, sensitive to the mean
redshifts of each redshift bins, and less to the statistical errors of
photo-$z$'s or the detailed shape of redshift distribution.  The level
of systematic photo-$z$ errors in each tomographic bins is quantified
in Figure~\ref{fig:bias_scatter}. The central values and error bars in
this plot show the bias in mean redshift and the statistical error of
the mean redshift, $\sigma(\langle \Delta z \rangle)$, where $\Delta
z$ is defined before (see around Fig.~\ref{fig:pdfscatter}) and the
average $\langle \cdots \rangle $ denotes the average over all the
galaxies in the tomographic redshift bin\footnote{The statistical
error of the mean redshift is reduced from the typical photo-$z$ error
of each galaxy as $\sigma(\langle z \rangle)\simeq \sigma(z_{\rm
ph})/{\rm N}$ with $N$ being the number of galaxies contained in the
redshift bin.  }.  Note that, for illustrative purpose, the error bars
are scaled for a survey area of $1$ arcmin$^2$, and therefore the
corresponding errors for our fiducial survey area of 2000 sq. degrees
are much smaller than plotted, by a factor of
$\sqrt{2000\times60^2}\simeq 2700$.

The middle- and bottom-row panels are the results of subsamples
obtained by further imposing the condition $0.2\le z_p\le 1.5$ or
$22.5\le i\le 25$, respectively.  It is clear that, without clipping
ill-defined photo-$z$ galaxies, the bias and errors are
significant. Note that the redshift bias for no tomography case
sometimes becomes smaller than in some tomographic bins (especially
highest redshift bins), because photo-$z$ outliers at low- and
high-redshifts cancel out to some extent in no tomography case.  As
can be seen from the middle-row panels, when the restricted redshift
range of $0.2<z<1.5$ is considered, a subsample with most accurate
photo-$z$'s is obtained, because spectral features of galaxies,
especially the Lyman and 4000 \AA~ breaks, are well captured by the
sets of filters in this redshift range.

\begin{figure*}[thpb]
  \begin{center}
   \epsscale{.8}
   \plotone{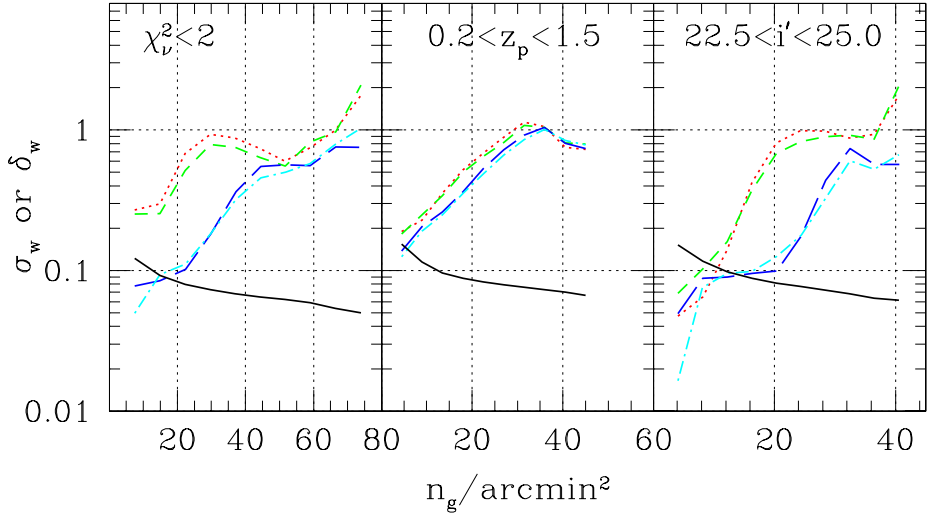}
   \caption{The forecasted constraints on the dark energy equation of
     state parameter $w$ as a function of number densities of galaxies
     included in the corresponding galaxy catalogs, expected for the
     lensing tomography experiment with survey area of $2000$~deg$^2$ in
     combination with the Planck CMB information. The three redshift bins
     are considered for each galaxy subsamples as in
     Figure~\ref{fig:nzsp}.  The solid curve in each panel shows the
     marginalized error $\sigma(w)$ assuming no photo-$z$ errors.  The
     other curves show the offset bias of the best-fit $w$ from the true
     value ($w=-1$), computed by using the Fisher matrix formalism (see
     around Eq.~[\ref{eq:bias_theta_red}]): the dotted, dashed,
     long-dashed and dot-dashed curves are for combinations of filters,
     $grizy$, $ugrizy$, $grizyJHK$ and $ugrizyJHK$, respectively.  The
     left, middle and right panels are the results for different object
     selections as in Figure~\ref{fig:nzsp}.
     \label{fig:bias1}
   }
  \end{center}
\end{figure*}

\begin{figure}[bhtp]
  \plotone{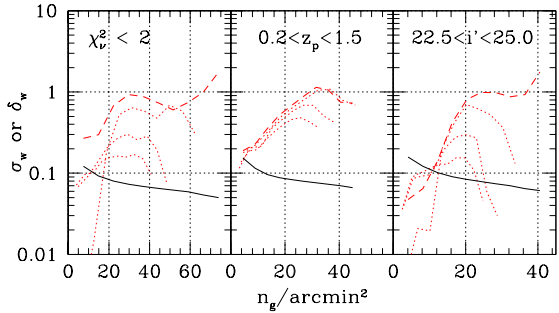}
  \caption{As in the previous figure, but the results obtained by
    artificially discarding photo-$z$ outliers given as
    $\log(1+z_p)/(1+z_s)\ge \pm 0.1$, $0.3$ and $0.5$ (the three dotted
    curves from top to bottom, respectively). The filter combination
    $grizy$ is considered in this plot. For comparison the top dashed
    curve in each panel is the same as the dotted curve in
    Figure~\ref{fig:bias1}.
    \label{fig:bias_out}
  }
\end{figure}

We next propagate the errors of tomographic redshifts into the dark
energy parameter $w$, expected from the lensing power spectrum
measurements, based on the Fisher matrix formalism described in
\S~\ref{ssec:fisher}.  To do this we address the following questions:
\begin{itemize}
\item{}
  The trade-off of dark energy constraint: the statistical accuracy
  of $w$, $\sigma(w)$, versus the offset of the best-fit value from the
  true value, $\delta w$.
\item{}
  Study the dark energy trade-off against different subsamples of
  galaxies.
\end{itemize}
These can be studied by using the simulated galaxy catalogs. The
statistical error of $w$ can be reduced by including more number of
galaxies in the sample for a fixed range of working multipoles ($5\le
\ell \le 3000$), because the shot noise is more suppressed. On the
other hand, a bias in the best-fit $w$ due to the photo-$z$ errors can
be reduced by discarding photo-$z$ outliers, leaving a fewer number of
galaxies in the subsample. Hence a trad-off point in $\sigma(w)$
versus $\delta w$ may be found by compromising these competing
effects.

Figure~\ref{fig:bias1} shows the marginalized error $\sigma(w)$ and
the amount of bias $|\delta w|$ as a function of number densities of
galaxies included in the corresponding galaxy subsamples. Again note
that we considered the lensing tomography with three tomographic bins
for a sky coverage of 2000 square degrees. The smaller number
densities in the horizontal axis correspond to subsamples of galaxies
where more galaxies with ill-defined photo-$z$'s are discarded by
imposing more stringent thresholds on ${\rm Var}(z)$, i.e. smaller
threshold values of ${\rm Var}(z)$, in the clipping method.  The
different curves are for different combinations of filters.  The error
$\sigma(w)$ is computed from the underlying true redshift
distribution, i.e. for the case with perfect photo-$z$'s, therefore
specified by the number density in the horizontal axis.  When $\delta
w\ge \sigma(w)$, the best-fit value of $w$ can be away from the true
one by more than the $1\sigma$ error; even if the true model has the
cosmological constant ($w= -1$), the result of $w\ne -1$ may be
falsely inferred. Hence a minimal requirement on photo-$z$ accuracies
can be assessed from the condition $\delta w\le \sigma(w)$.

First, the plot shows that, as the subsample is restricted to galaxies
with more accurate photo-$z$'s, i.e. the smaller number densities, the
bias in $w$ is reduced to some extent. On the other hand, the error
$\sigma(w)$ is only slightly degraded because the constraint comes
mainly from the sample variance limited regime for a given range of
working multipoles ($5\le l\le 3000$).

It is also shown that the bias can be reduced by adding the NIR-
and/or u-band data.  However, the broadband data alone may not be
sufficient to reduce the bias.  The optimal range of redshifts needs
to be considered, and a brighter subsample whose galaxies have higher
$S/N$ values in each filter is preferred to sufficiently reduce the
bias, as implied from the middle and right panels.  Depending on the
available set of filters, the compromising point can be obtained
around the number densities $\bar{n}_g=[10,30]$ arcmin$^{-2}$,
i.e. more than 60\% of ill-defined photo-$z$ galaxies need to be
discarded.  It would also be worth noting that combining the lensing
constraints with other dark energy probes such as the baryon acoustic
oscillation experiment may allow to further calibrate photo-$z$ errors
by breaking parameter degeneracies.

We have so far paid special attention to how to eliminate photo-$z$
outliers in order to obtain a subsample of galaxies suitable for
tomographic lensing measurements.  However, due to the limitation of
photo-$z$ accuracies, there may remain a residual bias in the
tomographic redshift bins, even if photo-$z$ outliers are completely
removed. To study this, Figure~\ref{fig:bias_out} shows the results
obtained by artificially discarding photo-$z$ outliers according to
the clipping criteria
\begin{equation}
  \log\frac{1+z_p}{1+z_s}>\pm t,
  \label{eq:outliers}
\end{equation}
with $t=0.1, 0.3 $ and $0.5$, respectively. The figure shows that a
bias in $w$ cannot be fully eliminated even if the outliers are
completely discarded. This implies that there remains a residual bias
in the mean redshift for each tomographic bins due to asymmetric
photo-$z$ errors around $z_p = z_s$, therefore the residual biases
would need to be calibrated, e.g. by using a spectroscopic training
subsample \citep[e.g.][]{MaBernstein:2008}.

%------------------------------------------------------------------------------
\subsection{Angular cross-correlations of galaxies between different
  photo-$z$ bins}
\label{ssec:ggcor}
%------------------------------------------------------------------------------
An alternative way to identify photo-$z$ outliers is using angular
cross-correlations of galaxies between different photo-$z$ bins
\citep[][]{Newman:2008,Erbenetal:2009,Zhang.etal:2009,Schulz:2009}. As
implied in Fig.~\ref{fig:nzsp}, photo-$z$ errors cause overlaps of
galaxies between different redshift bins. Therefore photo-$z$ errors
may cause non-vanishing cross-correlations of galaxies between
different photo-$z$ bins, if the galaxies indeed have similar true
redshifts, therefore are physically correlated with each other. In
other words the cross-correlations can, albeit statistical, be used to
monitor a contamination of photo-$z$ outliers. In this subsection we
use our simulated photo-$z$ catalogs to estimate the expected
signal-to-noise ratios for measuring the cross-correlations assuming
the same survey parameters we have considered.

Assuming the Limber approximation, the angular power spectra of
galaxies in the $i$- and $j$-th photo-$z$ bins are given as
\begin{equation}
  C^{gg}_{ij}(\ell)
  =
  \int_0^\infty\!\!dz \frac{dz}{d\chi} 
  \frac{b_i b_j}{\chi^2}
  \frac{n_i(z)n_j(z)}{\bar{n}_i \bar{n}_j}
  P_\delta\!\!\left( \frac{\ell}{\chi} , z \right)
  +
  \frac{\delta_{ij}^K}{\bar{n}_i},
\end{equation}
where $n_i(z)$ is the underlying true redshift distribution for the
$i$-th photo-$z$ bin and $\bar{n}_i$ is its mean number density.  In
the following we consider a sharp redshift binning in photo-$z$ space,
however, the underlying true distributions generally have overlaps due
to photo-$z$ errors.  We here simply assume that the galaxy
distribution in the $i$-th bin is related to the matter distribution
via constant bias parameter $b_i$, which is taken to $b_i=1$ for all
the photo-$z$ bins for simplicity.  To make this assumption
reasonable, we restrict the following analysis to a range of low
multipoles $l<500$.  Notice that the cross power spectra are not
affected by shot noise.

The strength of cross-correlations or redshift leakages can be
quantified by the cross-correlation coefficients at each multipole: 
\begin{equation}
  \mu_{ij}(\ell)
  =
  \frac{C_{ij}^{gg}(\ell)}{\sqrt{C_{ii}^{gg}(\ell)C_{jj}^{gg}(\ell)}}.
\label{eqn:mu_ij}
\end{equation}
The coefficient $\mu_{ij}\simeq 1$ implies significant
cross-correlations between the $i$- and $j$-th bins compared to their
auto-spectra, while $\mu_{ij}=0$ means no cross-correlation or no
leakage of photo-$z$ outliers into different bins. The total
signal-to-noise ratios expected for measuring the cross-correlations
can be estimated as
\begin{equation}
  \left(\frac{S}{N}\right)^2_{ij}
  \equiv
  \sum_{\ell=\ell_{\rm min}}^{\ell_{\rm max}}
  f_{\rm sky}(2\ell+1)
  \frac{C_{ij}^2}{ C_{ij}^2 + C_{ii} C_{jj} }.
\end{equation}
Here we simply assume the Gaussian covariances to model the
statistical errors in measuring cross power spectra from a
survey. Note that the error covariance includes the shot noise
contamination via the auto spectra $C_{ii}$ and $C_{jj}$. As for the
minimum and maximum multipoles used in the summation, we adopt
$\ell_{\rm min}=5$ and $\ell_{\rm max}=500$, respectively.

\begin{figure*}[t]
  \plotone{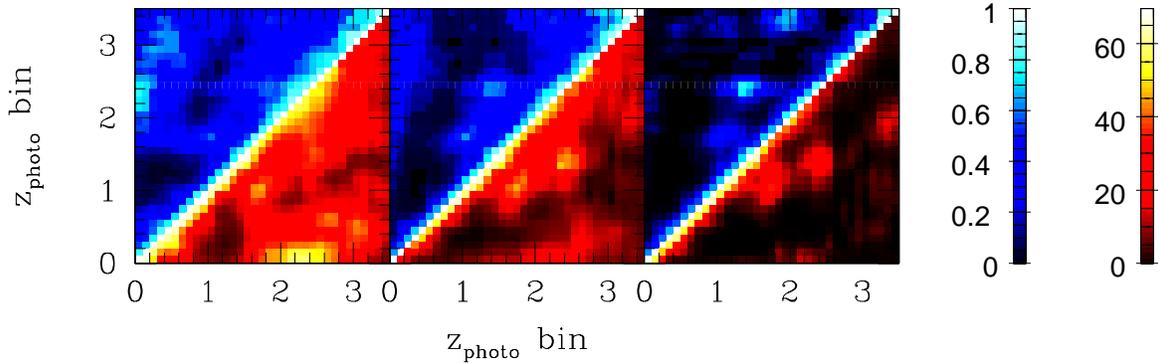}
  \caption{ 
    Angular cross-correlations between galaxies in two
    photo-$z$ bins, denoted on the horizontal and vertical axes, for
    our fiducial survey with 2000 square degrees, adopting 35 redshift
    bins over $0<z_p<3.5$ with the bin width $dz=0.1$.  As in
    Fig.~\ref{fig:scatter}, the left panel shows the result for galaxy
    catalog selected with $ugrizy$ and $\chi^2_\nu<2$, while the
    middle- and right-panels show the results for the catalogs where
    40\% and 70\% of galaxies with ill-defined photo-$z$'s are
    discarded.  The upper-left off-diagonal components in each panel
    show the correlation coefficients $\mu_{ij}$ (see
    Eq.~[\ref{eqn:mu_ij}]) at multipole $\ell =100$.  Compared with
    Fig.~\ref{fig:scatter}, one can find photo-$z$ outliers cause
    ``island'' regions with high coefficients $\mu_{ij}\simeq 1$. The
    lower-right components show the cumulative signal-to-noise ratios
    $(S/N)$ expected in measuring the cross-power spectrum over a
    range of multipoles $5 \le \ell\le 500$. Again the
    cross-correlations between different bins, caused by photo-$z$
    outliers, show high $S/N $ values greater than 10.
    \label{fig:ggcorr} 
  }
\end{figure*}

Figure~\ref{fig:ggcorr} studies the angular cross-correlations for our
fiducial survey parameters with 2000 square degree coverage.  Here we
consider the photo-$z$ galaxy catalog selected with $ugrizy$ and
$\chi^2_\nu<2$, and adopt 35 redshift bins over $0<z_p<3.5$ with the
bin width $dz=0.1$ corresponding to a typical photo-$z$ error on
individual galaxy basis.  The upper-left triangle in each panel shows
the correlation coefficients of cross-power spectra between the two
different photo-$z$ bins, $\mu_{ij}$. The coefficients have large
values around the diagonal terms, i.e. $z_{pi}\simeq z_{pj}$, because
the photo-$z$ errors cause significant overlaps between neighboring
redshift bins.  Compared with the results in Fig.~\ref{fig:scatter},
one can find that photo-$z$ outliers cause some isolated regions with
significant correlation coefficients.

The lower-right triangle shows the expected signal-to-noise ratios,
$S/N$, for measuring cross-correlations. The sufficiently high $S/N$
values, say greater than 10, can be expected for redshift bins that
have high correlation coefficients.  Thus monitoring the
cross-correlations between different photo-$z$ bins may allow to
further identify photo-$z$ outliers, in a statistical sense. However,
the genuine power of cross-correlation method for eliminating the
outliers or calibrating the photo-$z$ errors needs to be more
carefully studied.

Finally we remark on a more quantitative work done in
\cite{Schulz:2009}, which studied, based on mock simulations, the use of
cross-correlations of photometric galaxies with an overlapping
spectroscopic sample to calibrate the redshift distribution of the
photometric galaxies without using the photo-$z$ information. While the
main purpose of this paper is not using the cross-correlations to
identify photo-$z$ outliers, the promising result shown is that the
redshift distribution can be well reconstructed by using a sufficiently
{\em large} spectroscopic sample. However, one of the limiting factors
realized is the reconstruction requires a sufficiently fine binning of
spectroscopic redshifts, which tends to make the cross-correlation
measurements noisy. Therefore it would be interesting to study how the
method can be further refined by combining the cross-correlation method
and the photo-$z$ information, and the combined method may relax a
requirement on the size of spectroscopic calibration sample. We also
note that, as pointed out in \cite{BernsteinHuterer:2009}, the
cross-correlation method is affected by the lensing magnification bias,
which may cause an apparent correlations between foreground and
background galaxies even if there is no photo-$z$ errors to cause
redshift overlaps. This effect also needs to be included, and mock
simulations would be useful for such a study on the cross-correlation
method.

%%%%%%%%%%%%%%%%%%%%%%%%%%%%%%%%%%%%%%%%%%%%%%%%%%%%%%%%%%%%%%%%%%%%%%%%%%%%%%%
\section{Summary and Discussion}
\label{sec:conclusion}
%%%%%%%%%%%%%%%%%%%%%%%%%%%%%%%%%%%%%%%%%%%%%%%%%%%%%%%%%%%%%%%%%%%%%%%%%%%%%%%
In this paper we have studied how photo-$z$ errors available from
broadband multi-color data affect cosmological parameter estimation
obtained from tomographic lensing experiment. To do this, we made the
simulated mock galaxy catalog with photo-$z$ information constructed
from the COSMOS catalog.  Since the photo-$z$ errors are sensitive to
survey parameters such as available filters, the depths, and so on, we
considered in this paper the survey parameters to resemble the planned
Subaru Hyper Suprime-Cam survey, which is characterized by the optical
multi-passband data ($grizy$) and the depth $i\simlt 26$.  We also
studied how the photo-$z$ accuracy can be improved if combining the
optical data with the $u$-band data expected from a CFHT-type
telescope and the NIR ($JHK$) data from the VIKING-type
survey. However, the method developed in this paper can be readily
extended to other weak lensing surveys.

We particularly paid our attention to how to construct a galaxy
subsample suitable for weak lensing tomography. We showed that
photo-$z$ outliers can be efficiently identified by monitoring the
posterior likelihood of redshift estimation for each galaxy: more
exactly, the width of likelihood function defined by the second moment
around the best-fit redshift parameter (see Eq.~[\ref{eq:secmom}]) was
used as an indicator of the photo-$z$ accuracy on individual galaxy
basis.  It was also shown that the photo-$z$ outliers can be more
efficiently removed by restricting the ranges of working magnitudes
and/or redshifts, and adding the $u$- and $NIR$ bands (see
Figs.~\ref{fig:scatter}--\ref{fig:bias_scatter}).

Using the Fisher matrix formalism, we estimated how the photo-$z$
errors in a defined galaxy catalog cause biases in cosmological
parameters, especially the dark energy equation of state parameter
$w$. It was shown how the parameter biases can be reduced with
discarding galaxies with ill-defined photo-$z$ estimates. However,
with discarding more galaxies, the statistical accuracies of
parameters are degraded due to the increased shot noise
contamination. We found that the trade-off point, where the parameter
bias becomes similar or smaller than the marginalized statistical
error, can be achieved if a large fraction of ill-defined photo-$z$
galaxies ($\sim 70\%$) are discarded and if combined with the $u$- and
NIR-band data sets (Fig.~\ref{fig:bias1}).

However, as demonstrated in Fig.~\ref{fig:bias_out}, even if photo-$z$
outliers are completely eliminated, there may remain a non-negligible,
residual bias in the mean redshift of each tomographic bin because the
scatters around the relation between photometric and true redshifts,
$z_p=z_s$, are not necessarily symmetric and therefore not perfectly
canceled even after the average of galaxies in each redshift
bin. Therefore a careful calibration of the residual photo-$z$ errors
will be inevitably needed for any future surveys
% AJN
\citep{Hearin.etal:2010}.

A powerful method for the photo-$z$ calibration is using a training
spectroscopic subsample.  Naively, if a {\em fair, representative}
spectroscopic subsample of imaging galaxies used in the lensing
analysis is available, it allows a calibration of photo-$z$
errors. However, the size of such a spectroscopic sample required for
achieving the meaningful dark energy constraint becomes very large;
for a lensing survey with sky coverage of more than 1000 sq. degrees,
containing more than $10^8$ imaging galaxies, a subsample with more
than $ 10^6$ spectroscopic redshifts is required
\citep{Huterer.etal:2006,Ma.etal:2006}. Note that the currently
largest redshift sample is given by the COSMOS project containing
$10^4$ redshifts. Thus a survey collecting $10^6$ spectra, which is
required for our sample fully calibrated, is observationally very
expensive and almost infeasible, especially if redshifts of faint
galaxies are needed \citep[but see][for relaxing the requirement]
{BernsteinHuterer:2009}.

On the other hand, there is a new method recently proposed
\citep{Mandelbaumetal:2008,Limaetal:2008,Cunhaetal:2009}, using a
spectroscopic subsample of smaller size, which is not necessarily a
fair, representative subsample of imaging galaxies.  First
spectroscopic galaxies are compared to imaging galaxies in
multi-dimensional color space, rather than the photo-$z$ space.
Secondly the ratio between number densities of spectroscopic and
imaging galaxies is computed at each point in multi-color space. Then
the ratio is multiplied to the redshift distribution of spectroscopic
subsample to infer the underlying redshift distribution of imaging
galaxies. Thus this weighting method may allow the photo-$z$
calibration using a spectroscopic subsample of smaller size.  However,
there is still an open issue to be carefully investigated in this
method. For example, it is unclear how the calibration degrades if the
spectroscopic subsample has significant sample variance fluctuations
in the redshift distribution, e.g. due to clustering contamination at
particular redshifts due to a finite area coverage.

Another calibration method is using cross-correlations of galaxies in
photo-$z$ bins, as partly studied in Fig.~\ref{fig:ggcorr}. Again the
non-vanishing cross-correlations only arise when the photo-$z$ errors
cause leakages into different bins of true redshifts. Or spectroscopic
galaxies in the same survey region, if available, can also be used to
cross-correlate with imaging galaxies in order to calibrate the
photo-$z$ errors over a range of redshifts covered by the
spectroscopic sample \citep{Newman:2008,Schulz:2009}.  However
spectroscopic galaxies may be correlated with only particular types of
galaxies, therefore, this method may have a limitation.  Hence it
would be interesting to explore how to calibrate the redshift
distribution of imaging galaxies down to the required accuracy level
by combining various methods, the method developed in this paper and
the methods based on spectroscopic subsample or/and cross-correlation
measurements.

Finally we comment on another important contaminating effect, the
intrinsic alignment in galaxy shapes
\citep[e.g.][]{HirataSeljak:2004,Mandelbaumetal:2006,Mandelbaumetal:2009}.
As studied in detail in \cite{KingSchneider:2003} \citep[also
see][]{HeymansHeavens:2004,TakadaWhite:2004,BridleKing:2007}, accurate
photo-$z$ information is needed to calibrate and/or correct for the
intrinsic alignment contamination in weak lensing tomography. A
subsample with reliable photo-$z$'s, constructed based on the method
in this paper, may be also useful for this purpose.

%%%%%%%%%%%%%%%%%%%%%%%%%%%%%%%%%%%%%%%%%%%%%%%%%%%%%%%%%%%%%%%%%%%%%%%%%%%%%%%
\section*{Acknowledgments}
%%%%%%%%%%%%%%%%%%%%%%%%%%%%%%%%%%%%%%%%%%%%%%%%%%%%%%%%%%%%%%%%%%%%%%%%%%%%%%%
We would like to thank Rachel Mandelbaum and the members of the Hyper
Suprime Cam weak lensing working group for useful discussions and
comments.  We acknowledge the use of publicly available codes, {\it
HyperZ}, {\it LePhare} and CAMB.  This work is in part supported in
part by Japan Society for Promotion of Science (JSPS) Core-to-Core
Program ``International Research Network for Dark Energy'', by
Grant-in-Aid for Scientific Research from the JSPS Promotion of
Science (18072001,21740202), by Grant-in-Aid for Scientific Research
on Priority Areas No. 467 ``Probing the Dark Energy through an
Extremely Wide \& Deep Survey with Subaru Telescope'', and by World
Premier International Research Center Initiative (WPI Initiative),
MEXT, Japan.

%%%%%%%%%%%%%%%%%%%%%%%%%%%%%%%%%%%%%%%%%%%%%%%%%%%%%%%%%%%%%%%%%%%%%%%%%%%%%%%
%\bibliography{bibdata} 
%\bibliographystyle{apj}

\end{document}